\documentclass[12pt]{article}
\usepackage{epstopdf}
\usepackage{epsfig}
\usepackage{graphicx}
\usepackage{a4}
\usepackage{amsmath}
\usepackage{latexsym}
\usepackage{cite}

\usepackage{lineno}
\usepackage{color}
\usepackage{colordvi}

\graphicspath{{Pics/}}
\DeclareGraphicsExtensions{.eps,.ps}

\usepackage{pslatex}
\usepackage[latin1]{inputenc}
\usepackage[T1]{fontenc}

\newcommand{\msbar}{$\overline{\text{MS}}\, $}
\definecolor{cyan}{rgb}{0.6,0.3,0.0}

\begin{document}

\begin{titlepage}
\noindent
DESY 12-138 \hfill August 2012\\
LPN 12-085 \\
SFB/CPP-12-52 \\
\vspace{1.3cm}

\begin{center}
  {\bf 
\Large
Determination of the charm-quark mass in the $\overline{\text{MS}}$ scheme using charm production data 
from deep inelastic scattering at HERA.
  }
  \vspace{1.5cm}

  {\large
    S. Alekhin$^{\,a}$, 
    K. Daum$^{\,b,}$\footnote{Permanent address: DESY, 
    Notkestra{\ss}e 85, D--22607 Hamburg, Germany }, 
    K. Lipka$^{\,c}$
    and
    S. Moch$^{\,d}$
  }\\
  \vspace{1.2cm}

  {\it 
    $^a$Institute for High Energy Physics \\
    142281 Protvino, Moscow region, Russia\\
    \vspace{0.2cm}
    $^b$Bergische Universit{\"a}t Wuppertal \\ 
    Gau{\ss}stra{\ss}e 20, D-42097 Wuppertal, Germany \\
    \vspace{0.2cm}
    $^c$Deutsches Elektronensynchrotron DESY \\
    Notkestra{\ss}e 85 D--22607 Hamburg, Germany \\
    \vspace{0.2cm}
    $^d$Deutsches Elektronensynchrotron DESY \\
    Platanenallee 6, D--15738 Zeuthen, Germany \\
  }
  \vspace{2.4cm}

\large
{\bf Abstract}
\vspace{-0.2cm}
\end{center}
We determine the charm-quark mass $m_c(m_c)$ in the \msbar\ scheme using measurements of charm 
production in deep-inelastic $ep$ scattering at HERA in the kinematic range of photon virtuality 
$5$~GeV$^2<Q^2<1000$~GeV$^2$ and Bjorken scaling variable $10^{-5}<x<10^{-1}$. 
The extraction of $m_c$ from this process with 
space-like kinematics provides complementary information to results from hadronic processes. 
The QCD analysis of the HERA data yields a value 
of $m_c(m_c) = 1.27\pm 0.05 (\text{exp})^{+0.06}_{-0.01}(\text{scale})$ GeV 
at next-to-leading order and 
of $m_c(m_c) = 1.36\pm 0.04 (\text{exp})^{+0.04}_{-0.00}(\text{scale})\pm 0.1 (\text{theory})$ GeV 
at approximate next-to-next-to-leading order.
The results are consistent with and of comparable precision as the world average.
\vfill
\end{titlepage}

%
%
\newpage

\section{Introduction}
Quark masses are fundamental parameters of the Standard Model. However, 
due to confinement no free quarks are observed in nature. Therefore, for the determination of heavy quark masses  
a careful theoretical description is needed for appropriate observables.
This then enables the determination of heavy quark masses by comparing quark mass dependent theoretical predictions 
to experimental data. In such an analyses reference must be made to the specific scheme used to define the quark mass 
and it is mandatory to include radiative corrections in Quantum Chromodynamics (QCD) beyond the leading order (LO).

In QCD predictions for hard scattering processes employing the pole (or on-shell) scheme for the definition of the heavy-quark 
mass, $m_q$ is chosen to coincide with the pole of the heavy-quark propagator at each order in perturbative QCD. 
It is known since long, that the concept of pole masses in QCD has an inherent drawback, see, e.g.,~\cite{Bigi:1994em}.
Since quarks are confined inside hadrons, there is no pole in the quark propagator of the full theory.
As a consequence, the quark mass parameter in the pole mass scheme is limited to the perturbative domain with corrections of 
order ${\cal O}(\Lambda_{\text{QCD}}/m_q)$ and its value depends strongly on the order of perturbation theory.
Alternatively, QCD predictions for the charm  production cross section in DIS
can be considered as a function of the running mass $m_q(\mu_r)$ at the renormalisation 
scale $\mu_r$ in the \msbar\ scheme. The quark mass, $m_q(\mu_r)$, is treated on the same footing as the strong coupling 
constant $\alpha_s(\mu_r)$ and represents a prominent example of the so-called short-distance 
mass definition, which requires the heavy-quark masses to be evaluated at the scale $\mu_r$ 
much larger than the QCD scale $\Lambda_{\text{QCD}}$, i.e., $\mu_r \gg \Lambda_{\text{QCD}}$.

To this date, the charm-quark mass has been measured very accurately from data in
electron-positron annihilation, e.g., with the help of QCD sum rules, or
from numerical simulations on the lattice, see~\cite{Beringer:2012}.
These determinations are either entirely non-perturbative (lattice) or 
related to scattering processes in time-like kinematics only.
It is therefore of particular interest to examine the possibility of alternative charm-quark
mass determination from heavy-quark production in deep-inelastic scattering (DIS).
This reaction proceeds in the space-like kinematics with the distinct structure of the QCD amplitudes 
and in this way it provides an additional check of the QCD parameters' universality.

In this paper the theoretical framework developed in~\cite{Alekhin:2010sv} is applied 
to determine the charm-quark mass at next-to-leading order (NLO) and approximate 
next-to-next-to-leading order (NNLO). The recent results from the H1 
experiment~\cite{Aaron:2009jy,Aaron:2011gp,Aaron:2009af} on charm production in 
deep-inelastic $ep$ scattering at HERA are used in addition to the data sets of~\cite{Alekhin:2012ig}. 
By using the \msbar\ scheme the current 
analysis profits from improved stability of the perturbative expansion and the reduced theoretical 
uncertainty due to missing higher order contributions estimated by the variations of the renormalisation 
and factorisation scales. 

This paper is organised as follows: after a brief reminder of the theoretical basis for 
the treatment of heavy quarks in DIS we present the determination of the charm-quark mass in the \msbar\ scheme, 
based on recent charm production measurements from H1.
Special emphasis is put on the analysis procedure to account for the charm-quark mass dependence
of the $D^{*\pm}\,$ production cross section in the fiducial kinematic range. 
Finally, the running charm-quark mass $m_c(\mu_r)$ is determined at NLO and at
NNLO in QCD. In the NNLO case, 
the theory predictions are available only to an approximation with a substantial uncertainty arising from missing 
information in the NNLO DIS Wilson coefficients. The results are compared to previous extractions 
of $m_c(m_c)$ from the DIS data in~\cite{Alekhin:2010sv,Alekhin:2011jq} and to the results of different 
determination methods, used for the world average~\cite{Beringer:2012}.

\section{Theory framework}
\label{sec:theory}
Charm-quark production in neutral-current DIS proceeds by scattering of a charged lepton $e$ off a 
proton $P$ following the reaction
\begin{equation}
  \label{eq:hqborn}
  e(l) \,+\, P(p) \:\:\rightarrow\;\: e(l^{\,\prime}) \,+\, c^{}(p_1^{}) \,+\, \bar{c}^{}(p_2^{}) \,+\, X \; ,
\end{equation}
in which a virtual boson of space-like momentum $Q^2=-q^2=-(l-l^{\,\prime})^2$ is exchanged. 
Here $l$, $p$, $l^\prime$, $p_1$ and $p_2$ denote the four-vectors of the incoming electron and proton and the outgoing electron,   charm quark and anti-charm quark, respectively.
The charm-quark pair, $c^{}\bar{c}^{}$, in the final state is heavy, such that $m_c^2 \gg \Lambda^2_{\text{QCD}}$ holds. 
We restrict the momentum transfer to be much smaller than the $Z$-boson mass $Q^2\ll M_Z^{\,2}$.

The inclusive cross section is expressed in terms of the standard DIS proton structure functions $F_k$ with $k=2,L$. 
The charm contributions to the inclusive structure functions $F_k$ are denoted by $F_k^{c\bar{c}}(x,Q^2,m_c^2)$.
In referring to reaction~(\ref{eq:hqborn}) the structure function $F_k^{c\bar{c}}$ 
requires by definition a charm-quark pair in the final state, but it is otherwise a completely inclusive quantity, 
especially with respect to the proton initial state. Thus, the cross section for the reaction~(\ref{eq:hqborn}) reads
\begin{eqnarray}
  \label{eq:cross-section}
  \frac{d^2\sigma^{c\bar{c}}}{dx\,dQ^2} \;= \;
  \frac{2\:\!\pi\,\alpha^2}{x\, Q^{\,4}} 
  \left[\left\{1+(1-y)^2\right\}\, F_2^{c\bar{c}}(x,Q^2,m_c^2) - y^2 F_L^{c\bar{c}}(x,Q^2,m_c^2)\right] 
  \, .
\end{eqnarray}
Here $\alpha$ is the electromagnetic coupling constant 
and the DIS variables $x$ and $y$ are defined as $x=Q^2/(2p\cdot q)$ 
and $y=(p\cdot q)/(p\cdot l)$, respectively.
For later use the reduced cross section is introduced as

\begin{eqnarray}
  \label{eq:red-crs}
 \sigma^{c\bar{c}}_{\rm red}(x,Q^2,m_c^2)  \;= \;
  F_2^{c\bar{c}}(x,Q^2,m_c^2) - \frac{y^2}{1+(1-y)^2}\, F_L^{c\bar{c}}(x,Q^2,m_c^2)
  \, .
\end{eqnarray}

In the standard factorisation approach to perturbative QCD the structure functions 
$F_k^{c\bar{c}}$ can be written as a convolution of parton distribution functions (PDFs) 
and Wilson coefficients, see e.g.,~\cite{Laenen:1992zk,Riemersma:1994hv},
\begin{equation}
  \label{eq:totalF2c}
  F_k^{c\bar{c}}(x,Q^2,m_c^2) \;=\;
  {\alpha_s\, e_c^{\:\!2}\, Q^2 \over 4\:\! \pi^{\:\!2}\, m_c^2} \,\,
  \sum\limits_{i \,=\, q,{\bar{q}},g} \,\,
  \int_{x}^{\,z^{\,\rm max}}
      {dz \over z} \: f_{i}^{}\left({x \over z},\, \mu_f^2 \right)\,
      C_{k, i}^{}\left(z,\,\xi,\,\mu_f^2,\,\mu_r^2 \right)
      \; ,
\end{equation}
where $z^{\,\rm max} = 1/(1 + 4\:\! m_c^2/Q^2)$ 
and $e_c^{}=2/3$ is the normalised charm-quark charge. 
The PDFs for the parton of flavor $i$ are denoted as $f_{i}(x,\mu^2_f)$ 
and the sum in equation~(\ref{eq:totalF2c}) runs over all flavor combinations, 
i.e., singlet and non-singlet, and the gluon.
The Wilson coefficients $C_{k,i}$ depend on the kinematic variables $z$ and $\xi$,  
\begin{equation}
  \label{eq:s-xi-def} 
  z \;=\;{Q^2 \over Q^2+s}
  \, ,
  \qquad
  \xi \;=\; {Q^2 \over m_c^2} 
  \, ,
\end{equation}
with $s$ denoting the  partonic centre-of-mass energy.

For the treatment of heavy quarks in DIS the so-called fixed flavour number scheme (FFNS) 
is chosen with the number of active quarks in the proton $n_f = 3$ and massive charm quarks 
appearing exclusively in the final state. 
Moreover, the strong coupling constant is defined in the same scheme as $\alpha_s(n_f)$ 
with $n_f = 3$ for charm quark production.
This description of QCD with one massive and $n_f$ light quarks can be related to QCD 
with ($n_f+\!1$) light quarks by means of the standard matching conditions, cf. 
e.g.,~\cite{Buza:1995ie,Bierenbaum:2007qe}, and for a discussion of all-order resummations of logarithms 
in $Q^2/m_c^2$ in so-called variable flavor number schemes (VFNS)~\cite{Buza:1996wv}, 
see~\cite{Alekhin:2009ni,Forte:2010ta}.
The $C_{k,i}$ are obtained in perturbative QCD as an expansion in $\alpha_s = \alpha_s(\mu_r)$,
\begin{eqnarray}
  \label{eq:coeff-exp}
  C_{k, i}^{}(z,\xi,\mu^2)
  \; = \;
  \sum\limits_{j=0}^{\infty}\, (4\:\! \pi\, \alpha_s)^j \, 
  c^{\,(j)}_{k,i}(z, \xi, \mu^2)
  \; = \;
  \sum\limits_{j=0}^{\infty}\, (4\:\! \pi\, \alpha_s)^j \,
  \sum\limits_{l=0}^{j} c^{\,(j,\ell)}_{k,i}(z, \xi)\: 
      \ln^{\,\ell}\frac{\mu^2}{m_c^2} \:\; ,\quad
\end{eqnarray}
here the renormalisation and factorisation scales are set to $\mu\,=\,\mu_f\,=\,\mu_r$. In order 
to estimate the uncertainties due to missing higher orders, the scale $\mu$ is varied by a factor 
two up and down around the nominal value
\footnote{Most of current global PDF determinations assume $\mu_r=\mu_f = Q$ in fits 
to DIS data. This is the appropriate scale choice for massless structure functions and in 
general for large values of $Q$ when mass effects are negligible.}
\begin{eqnarray}
  \label{eq:scalevar}
  \mu\, \in\, \kappa \sqrt{Q^2+4 m_c^2}
  \,
  \quad\quad 
  \text{with} 
  \quad\quad \frac{1}{2} \le \kappa \le 2.
\end{eqnarray}

Conventionally, the Wilson coefficients are presented using the pole mass scheme for $m_c$, see 
e.g.,~\cite{Laenen:1992zk,Kawamura:2012cr}. The necessary formulae for the conversion to the running 
mass $m_c(\mu_r)$ in the \msbar\ scheme are well known~\cite{Gray:1990yh,Chetyrkin:1999qi,Melnikov:2000qh} 
and the application of the \msbar scheme to the calculation of $F_2^{c\bar{c}}$ has been detailed in~\cite{Alekhin:2010sv} 
(see also~\cite{Langenfeld:2009wd,Aliev:2010zk} for related work on heavy-quark hadro-production).
For the inclusive cross sections at short distances an appropriate choice for the scale 
of the running mass $m_c(\mu_r)$ is $\mu_r=m_c$.
The renormalisation group evolution for the scale dependence governed by the 
corresponding quark mass anomalous dimension~\cite{Vermaseren:1997fq,Chetyrkin:1997dh}
converges even for scales as low as the charm-quark mass, 
that is ${\cal O}(1)$~GeV.

The Wilson coefficients in equation~(\ref{eq:coeff-exp}) have been computed exactly to NLO
and the functions $c^{\,(1)}_{k,i}$~\cite{Laenen:1992zk} are often used via the parameterisations
of~\cite{Riemersma:1994hv}, see also~\cite{Harris:1995tu}.
At NNLO approximate results for the 
most important gluon and quark coefficient functions,  
$c^{\,(2)}_{2,g}$ and $c^{\,(2)}_{2,q}$, are known~\cite{Kawamura:2012cr} 
and denoted by NNLO$_\text{approx}$ in the following.
These are based on recent partial NNLO improvements for the structure function $F_2^{c\bar{c}}$ 
which encompass various kinematic limits:
\begin{enumerate}
\item[(i)] $c_{2,g}^{\,(2)\,{\rm thr}}$ near the partonic threshold for $s \simeq 4 m_c^2$\, , 
\item[(ii)] $c_{2,g}^{\,(2)\,{\rm asm}}$ and $c_{2,q}^{\,(2)\,{\rm asm}}$ 
  at asymptotically high scales $Q^2 \gg m_c^2$\, , 
\item[(iii)] $c_{2,g}^{\,(2)\,{\rm small-x}}$ and $c_{2,q}^{\,(2)\,{\rm small-x}}$ 
  at high energies $s \gg m_c^2$ (small-$x$).
\end{enumerate}
The threshold approximation $c_{2,g}^{\,(2)\,{\rm thr}}$ has been determined 
to the next-to-next-to-leading logarithm (NNLL)
(see~\cite{Laenen:1998kp,Alekhin:2008hc} for previous approximations at the
level of the next-to-leading logarithm (NLL)).
The function $c_{2,q}^{\,(2)\,{\rm thr}}$ is consistent with zero to the accuracy considered.
Likewise, fully analytic results for $c_{2,i}^{\,(2)\,{\rm asm}}$, $i=g,q$, 
in the asymptotic regime of $Q^2 \gg m_c^2$ have been obtained.
The corresponding calculations make use of the formalism of~\cite{Buza:1995ie,Bierenbaum:2007qe} 
and a number of lowest even-integer Mellin moments for the necessary heavy-quark operator matrix elements 
at three loops~\cite{Bierenbaum:2008yu,Bierenbaum:2009mv,Bierenbaum:2009zt}, see also~\cite{Ablinger:2010ty}.
In the high-energy limit the expression for $c_{2,i}^{\,(2)\,{\rm small-x}}$ 
is exact to leading-logarithmic (LL) accuracy at small-$x$ due to~\cite{Catani:1990eg} but is approximate only at NLL.

The combination of all available information leads to an expression for the NNLO Wilson coefficients, $c_{2,i}^{(2)}$ of the form:
\begin{eqnarray}
  \label{eq:assembly}
  c_{2,i}^{\,(2)} &\!\simeq\!& 
  c_{2,i}^{\,(2){\,\rm thr}}
  \,+\, \left(1 - f(\xi) \right) c_{2,i}^{(2)\,{\rm asm}}
  \,+\, f(\xi)\, c_{2,i}^{(2)\,{\rm small-x}}
  \, ,
\end{eqnarray}
with a suitable matching function $f(\xi)$ connecting the regions $Q^2 \gg m_c^2$ and $s \gg m_c^2$, cf. equation~(4.9) in~\cite{Kawamura:2012cr}.

The approach chosen has one caveat, though, which is a non-negligible theoretical 
uncertainty due to the poorly constrained NLL term at small-$x$ in $c_{2,i}^{\,(2)\,{\rm small-x}}$ and  the limited number of known three-loop Mellin 
moments~\cite{Bierenbaum:2008yu,Bierenbaum:2009mv,Bierenbaum:2009zt}.
To account for these deficits two different scenarios $c_{2,i}^{\,(2),A}$ and $c_{2,i}^{\,(2),B}$ in equation~(\ref{eq:assembly}), carefully designed to parametrise  the related 
uncertainties, will be considered in the subsequent analysis.
For the exact definitions of $c_{2,i}^{\,(2),A}$ and $c_{2,i}^{\,(2),B}$ see equations~(4.17),~(4.18) and equations~(4.21),~(4.22) in~\cite{Kawamura:2012cr}. 
An additional theoretical uncertainty on $F_2^{c\bar{c}}\,$ is estimated by
the variation of the renormalisation and factorisation scales.
This part of the theoretical uncertainty is obtained using exact results since the $\mu$-dependence is fully known at NNLO~\cite{Laenen:1998kp,Alekhin:2010sv} from the renormalisation group, see also~\cite{Kawamura:2012cr}.

\section{QCD analysis of charm production measurements}
\label{sec:analysis}
To study the impact of the charm production measurements at HERA on the determination of the \msbar\ charm mass a variant of the ABM11 fit~\cite{Alekhin:2012ig} 
is performed with the data of~\cite{Aaron:2009jy,Aaron:2011gp,Aaron:2009af} included. 
The data 
were collected by the H1 experiment during the HERA II running period corresponding to an integrated luminosity of about 350 pb$^{-1}$ and cover the
kinematic range of photon virtualities $5$~GeV$^2<Q^2<2000$~GeV$^2$.
The charm mass $m_c(m_c)$ is a free parameter of the fit comparing equation~(\ref{eq:red-crs}) to the experimental data. 
Experimentally charm-quark production in DIS is tagged via fully reconstructed charmed mesons or by using secondary vertex information of tracks, exploiting the longevity and the large mass of charmed hadrons. These measurements are restricted (e.g. in the transverse momenta or angles of the produced particles) by the 
acceptance of the detector. The phase space, in which charmed hadrons can be fully reconstructed is usually referred to as 
the visible or fiducial phase space. Corrections for the non-measurable phase space and the fragmentation of charm quarks to charmed hadrons are applied in the fit. 
In the following, the measurements of charm production used as input and their treatment in the QCD analysis are described with particular 
emphasis on the corrections for non-measurable phase space.

The $c$-quark production in~\cite{Aaron:2009jy,Aaron:2011gp} is tagged via fully reconstructed  $D^{*\pm}$-mesons in the decay mode
$ D^{*\pm} \rightarrow (D^0\to K^\mp \pi^\pm)\pi^\pm$. These measurements are restricted to the visible kinematic range of the $D^{*\pm}$-meson's 
transverse momentum $p_T(D^*)>1.25$~GeV and pseudo-rapidities $|\eta(D^*)|<1.8$ at medium virtualities $Q^2<100$~GeV$^2$. 
At high virtualities $100$~GeV$^2$ $<Q^2<1000$~GeV$^2$ the visible range of $p_T(D^*)>1.5$ GeV and $|\eta(D^*)|<1.5$ is covered. 
Due to the phase space limitations equation~(\ref{eq:cross-section}) cannot be used directly in the analysis of these data. Therefore, for these measurements 
the invisible phase space region is accounted for in equation~(\ref{eq:cross-section}) through a factor 
 \begin{eqnarray}
  \label{eq:eps}
  \epsilon_{\rm vis} \,=\, \sigma_{\rm vis}(D^{*\pm})/\sigma^{c\bar{c}}_{\rm full}
  \, ,
\end{eqnarray}
where $\sigma_{\rm{vis}}(D^{*\pm})$ and $\sigma^{c\bar{c}}_{\rm full}$ are the QCD predictions for the $D^{*\pm}$ meson cross section in the visible phase 
space and the charm production cross sections in the full phase space, respectively. These predictions are calculated at NLO in the FFNS with the 
fully exclusive program HVQDIS~\cite{Harris:1995tu}. The contribution from $b$-quarks to the inclusive $D^{*\pm}\,$-meson production cross section, 
reported in~\cite{Aaron:2009jy,Aaron:2011gp}, is estimated using HVQDIS 
and is subtracted. The $D^{*\pm}\,$ cross sections are re-calculated using 
the recent branching ratio values~\cite{Beringer:2012}.

The value of $\epsilon_{\rm vis}$ depends strongly on $m_c$ and its dependence on  $m_c$ is taken into account in the fit.
For this purpose $\epsilon_{\rm vis}$ is calculated in a first step for selected values of the charm quark mass, which 
correspond to the range of $m_c(m_c)$ scanned in our fit:
\begin{equation}
  \label{eq:mcrun}
  m_c(m_c) \,=\, 
  0.9, 1.05, 1.2, 1.35~{\rm GeV}
  \, .
\end{equation}
Since HVQDIS is based on the pole-mass definition, the calculations are performed using the following values for the charm quark mass in pole definition:
\begin{equation}
  \label{eq:mcpole}
  m_c^{\rm pole} \,=\, 
  1.41, 1.53, 1.67, 1.81~{\rm GeV}
  \, ,
\end{equation}
obtained by mapping of the running-mass grid in equation~(\ref{eq:mcrun}) 
according to the matching between pole and \msbar\ definitions at the
appropriate order as detailed below.
The relation is known to $O(\alpha_s^3)$~\cite{Gray:1990yh,Chetyrkin:1999qi,Melnikov:2000qh},
\begin{equation}
  \label{eq:match}
  m_c^{\rm pole} \,=\, 
  m_c(m_c)\left[
  1+\frac{4}{3}a+ \left(-3.1242+13.4434 \right)a^2+116.504a^3
  \right]
  \ ,
\end{equation}
where $a=\alpha_s(m_c(m_c))/\pi$. 

In the calculation of $\epsilon_{\rm vis}$, the proton structure is described by special PDF sets, 
provided for this analysis. They correspond to the NLO variant of ABM11 fit~\cite{Alekhin:2012ig} and are performed in the \msbar 
definition of charm quark mass. The PDFs depend on the value of $m_c$ through substantial correlations between
$\alpha_s$, $m_c$ and the gluon PDF. In order to provide a fully consistent treatment of the charm-mass effects 
in $\epsilon_{\rm vis}$ one has to take into account this dependence. 
Therefore we use the PDFs, which exactly correspond to the current value of
$m_c$ appearing in the fit. 
These PDFs are given by interpolation 
between those obtained in the variants of ABM fit with the $m_c$ settings of equation~(\ref{eq:mcrun}).
The full dependence of $\epsilon_{\rm vis}$ on $m_c$ is provided by parabolic interpolation of the HVQDIS results obtained at the 
values of $m_c$ in equation~(\ref{eq:mcpole}) with the interpolation coefficients $P$, calculated independently for the LO and NLO terms in $\sigma^{c\bar{c}}_{\rm vis/\rm full}$,
\begin{eqnarray}
  \label{eq:hvlo}
  \sigma^{c\bar{c},\rm LO}_{\rm vis/\rm full}(x,Q^2,y,m_c^{\rm pole}) &=&
  \alpha_s \sum\limits_{i=0}^2 \left[P^{\rm vis/\rm full}_{0,i}(x,Q^2,y)(m_c^{\rm pole})^i\right]
  \, , \\
  \label{eq:hvnlo}
  \sigma^{c\bar{c},\rm NLO}_{\rm vis/\rm full}(x,Q^2,y,m_c^{\rm pole}) &=&
  \alpha_s^2 \sum\limits_{i=0}^2 \left[P^{\rm vis/\rm full}_{1,i}(x,Q^2,y)(m_c^{\rm pole})^i\right]
  \, . 
\end{eqnarray}
Such a representation allows the determination of $\epsilon_{\rm vis}$ in terms of the 
\msbar\ mass. This is achieved by 
substituting $m_c^{\rm pole}$ in equations~(\ref{eq:hvlo}) 
and (\ref{eq:hvnlo}) with the matching condition of equation~(\ref{eq:match}) 
similarly to the approach used earlier to derive the heavy-quark 
electro-production coefficient functions in the \msbar\ mass 
definition~\cite{Alekhin:2010sv}. 
The terms of $O(\alpha_s^3)$ and higher appearing after substitution exceed the NLO accuracy therefore they 
are dropped and the final expressions for $\sigma^{c\bar{c}}_{\rm vis/\rm full}$ employed in our analysis read
\begin{eqnarray}
\sigma_{cc,LO}^{\rm vis/\rm full}(x,Q^2,y,m_c(m_c)) 
&=&
\alpha_s \Sigma_{i=0}^2 \left[\bar P^{\rm vis/\rm full}_{0,i}(x,Q^2,y)(m_c(m_c))^i\right]
\, ,
\label{eq:hvlorun}
\\
\sigma_{cc,NLO}^{\rm vis/\rm full}(x,Q^2,y,m_c(m_c))
&=&
\alpha_s^2 \Sigma_{i=0}^2 \left[\bar P^{\rm vis/\rm full}_{1,i}(x,Q^2,y)(m_c(m_c))^i\right]
\, ,
\label{eq:hvnlorun}
\end{eqnarray}
with 
\begin{eqnarray}
\bar P^{\rm vis/\rm full}_{1,0} &=& P^{\rm vis/\rm full}_{1,0}
\, ,
\label{eq:crun0}
\\
\bar P^{\rm vis/\rm full}_{1,1} &=& P^{\rm vis/\rm full}_{1,1}+\frac{4a}{3\alpha_s}P^{\rm vis/\rm full}_{0,1}
\label{eq:crun1}
\, ,
\\
\bar P^{\rm vis/\rm full}_{1,2} &=& P^{\rm vis/\rm full}_{1,2}+\frac{8a}{3\alpha_s}P^{\rm vis/\rm full}_{0,2}
\label{eq:crun2}
\, ,
\end{eqnarray}
and
\begin{equation}
\bar P^{\rm vis/\rm full}_{0,i} \,=\, P^{\rm vis/\rm full}_{0,i}
\, ,
\label{eq:crun3}
\end{equation}
for $i=0,1,2$.

The cross section predictions $\sigma_{\rm vis}(D^{*\pm})$ depend not only on the kinematics of the charm quark 
production mechanism but also on the fragmentation of the charm quark into $D^{*\pm}$ mesons. 
The charm quark fragmentation function to $D^{*\pm}$ mesons has been measured by H1~\cite{h1frag} using the production 
of $D^*$-mesons with and without associated jets in DIS. In the calculation of $\sigma_{\rm vis}(D^{*\pm})$  
the longitudinal fragmentation is performed in the $\gamma^*-p$ centre-of-mass frame, 
using the fragmentation function of Kartvelishvili et al.~\cite{Kartvelishvili:1977pi} which is controlled by 
a single parameter, $\alpha_K$. This parameter has been determined
for two different regions of the partonic centre-of-mass energy squared, $s$, depending on the jet requirements 
made in the different analyses. The fragmentation is observed to become softer with increasing $s$ as
expected from the perturbative evolution of the fragmentation function not implemented in HVQDIS.
This is accounted for in the current analysis by using different values of $\alpha_K$ corresponding to the measurements 
for two ranges of $s$, as listed in table~\ref{tab:kart}. The limits on the ranges in $s$
are determined with HVQDIS by applying the jet requirements of the individual analysis on parton 
level. The $\alpha_K$ parameters and the $s$ ranges are varied according to their uncertainties 
to evaluate the corresponding uncertainty on  $\sigma_{\rm vis}^{c\bar{c}}$.
\begin{table}[h]
\begin{center}
\begin{tabular}{|c|c|l|} \hline
  $s$ range [GeV$^2$]    & $\alpha_K$    & measurement \\ \hline
  $s<(70\pm 40)$         & $6.1 \pm 0.9$ & \cite{h1frag} $D^*$, DIS, no-jet sample \\ \hline
  $(70 \pm 40) < s $ & $3.3 \pm 0.4$ & \cite{h1frag} $D^*$, DIS, jet sample \\ \hline
\end{tabular}
\end{center}
\caption{
  \label{tab:kart}
  \small
  The parameter $\alpha_K$ of the Kartvelishvili et al.
  fragmentation function for the $D^*$ mesons employed in our analysis 
  versus parton centre-of-mass energy squared $s$, equation~(\ref{eq:s-xi-def}). 
}
\end{table}
The charmed hadrons also receive a 
transverse momentum, $\bar{k}_T$, with respect to the charm quark direction according to
\begin{equation}
f(\bar{k}_T) = \bar{k}_T \exp(-2 \bar{k}_T/\langle \bar{k}_T\rangle)
\, . 
\end{equation}
The transverse momentum average $\bar{k}_T$ is chosen as $0.35\pm 0.15$~GeV$^2$, in line with the
experimental results on hadron production in $e^+e^-$ 
collisions~\cite{ptkink_aleph,ptkink_delphi,ptkink_markII,ptkink_pluto1,ptkink_pluto2,ptkink_tasso}.
A fragmentation fraction of  charm quarks into $D^{*\pm}\,$  mesons of $f(c\rightarrow D^{*+})=0.2287 \pm 0.0056$, is used as 
determined by averaging the $e^+e^-$ and $ep$ results~\cite{lohrmann}.
In order to evaluate the fragmentation model uncertainty on $\sigma_{\rm vis}(D^{*\pm}$)) the parameters $\bar{k}_T$, $\alpha_K$, and $f(c\rightarrow D^{*+})$ 
are varied within the uncertainties quoted above and each variation is considered as a correlated uncertainty in $\epsilon_{\rm vis}$.
In a few cases, a symmetric variation of the model parameters results in an asymmetric uncertainty on the cross section. In such cases, the largest 
absolute value of the uncertainty is assigned. The dominant systematic uncertainty is arising from the variation of the fragmentation function parameter $\alpha_K$. 

The measurement~\cite{Aaron:2009af} used in the fit is based on determination of the vertex displacement of the tracks and covers essentially the full phase space. 
However, the extrapolation to the cross sections of charm and beauty quark
production is performed with a Monte-Carlo simulation and the determination of a 
correction at NLO as in case of $D^{*\pm}$ analysis is not possible. We assume this correction to be small and no extrapolation factor is applied to these data.

\section{Results}
The NLO variant of our analysis applies $O(\alpha_s^2)$ corrections to the heavy quark electro-production Wilson 
coefficients~\cite{Laenen:1992zk} and the NLO ABM11 PDFs. For the NNLO variant of our analysis we use the NNLO ABM11 PDFs and 
apply the corrections up to $O(\alpha_s^3)$ as discussed in section~\ref{sec:theory}. Since the NNLO Wilson coefficients $c_{2,i}^{\,(2)}$ 
in equation~(\ref{eq:assembly}) are still approximate and affected by a residual uncertainty parameterised  by  $c_{2,i}^{\,(2),A}$ and $c_{2,i}^{\,(2),B}$ 
we define a particular shape for $c_{2,i}^{\,(2)}$ as a linear combination of the two envelopes with a parameter $d_N$ 
interpolating between these two options,
\begin{equation}
  c_{2,i}^{\,(2)} \,=\, 
  (1-d_N) c_{2,i}^{\,(2),A} + d_N c_{2,i}^{\,(2),B} 
  \, .
\end{equation}
The consistency of the prediction using results of the fit with the data 
of~\cite{Aaron:2009jy,Aaron:2011gp} on $\sigma_{\rm red}^{c\bar{c},{\rm VIS}}= \sigma_{\rm{vis}}(D^{*\pm})/f(c \rightarrow D^{*\pm})$ 
and of~\cite{Aaron:2009af} on $\sigma^{c\bar{c}}_{\rm red}$
is illustrated in figures~\ref{fig:h1sig} and \ref{fig:h1f2}, respectively. 
In the case 
of NLO, the fit quality is very good for each data set with the total value of $\chi^2=60$ for the number of data points (NDP) equal to 60. 
The consistency of the NNLO prediction using the fit results with the data is equally good as for the NLO variant of the fit 
(cf. figures~\ref{fig:h1sig} and~\ref{fig:h1f2}), with the best fit value of $\chi^2/NDP=63/60$ being achieved for $d_N=-0.6$. 
The variant of the NNLO Wilson coefficient given by $c_{2,i}^{\,(2),B}$ (equations~(4.18) and (4.22) in~\cite{Kawamura:2012cr}) 
is clearly disfavoured by the data with $\chi^2=156$, while 
the description provided by the variant 
$c_{2,i}^{\,(2),A}$ (equations~(4.17) and (4.21) in~\cite{Kawamura:2012cr}) with $\chi^2=72$ is comparable to the best one. 
Therefore we assign to the value of $m_c(m_c)$ at NNLO a conservative theoretical uncertainty of $100$~MeV which accounts 
for the incomplete current knowledge of the NNLO Wilson coefficient and which corresponds to the difference between the 
values of $m_c(m_c)$ obtained with $d_N=-0.6$ and $d_N=0$. 

As a result of the QCD analysis, we have determined the \msbar\ charm quark mass at NLO and at approximate NNLO accuracy in QCD, 
\begin{eqnarray}
  \label{eq:mcres-nlo}
  m_c(m_c) \,\,=&
  1.27\, \pm 0.05 (\text{exp})\,^{+0.06}_{-0.01} (\text{scale})
  \hspace*{30mm}
  &{\rm NLO}
  \, ,
  \\
  \label{eq:mcres-nnlo}
  m_c(m_c) \,\,=&
  1.36\, \pm 0.04 (\text{exp})\,^{+0.04}_{-0.00} (\text{scale})\, \pm 0.10 (\text{th})
  \hspace*{11mm}
  &{\rm NNLO_\text{approx}}
  \, .
\end{eqnarray}
The experimental uncertainty is obtained from the propagation of all
uncertainties in the data with account of their correlations. 
The fragmentation model uncertainties relevant for the data~\cite{Aaron:2009jy,Aaron:2011gp} are also included on the same footing.
To estimate the influence of the PDF uncertainty on the precision of the $m_c(m_c)$ determination, 
the NLO analysis is repeated for each of the ABM11 PDF set members, representing the $1\sigma$ 
uncertainty in the fitted PDF parameters\footnote{
The parameterisation of the proton structure functions in the ABM11 PDF set~\cite{Alekhin:2012ig} incorporates 
the effect from higher twist terms described by operators of dimension six in the framework 
of Wilson's operator product expansion.}.
The differences between the values $m_c(m_c)$ obtained in this way and the ones obtained with 
the central PDF member are added in quadratures. The resulting uncertainty 
on $m_c(m_c)$ is about $10$~MeV which is very small in comparison to the other uncertainties.

As the fit is performed at fixed order in perturbation theory, 
one needs to account for missing contributions beyond the order considered. 
This uncertainty is calculated in the standard manner from 
the variation of the renormalisation and factorisation scales within the range given in equation~(\ref{eq:scalevar}). 
The scale variation is performed simultaneously in equation~(\ref{eq:red-crs}) and in the calculation of the
extrapolation factors $\epsilon_{\rm vis}$. For the NNLO case the parameter $d_N$ is fixed at the value of $-0.6$, which is preferred by the fit. 
Due to the sensitivity of $\epsilon_{\rm vis}$ on the scale variation, 
the resulting uncertainty on $m_c(m_c)$ does not improve significantly 
from the NLO prediction to the NNLO one. In case of the NNLO result, the theory error on $m_c(m_c)$
of $100$~MeV due to lacking knowledge on the NNLO Wilson coefficient 
is significantly larger than the uncertainty due to scale variation. 
The necessary theory computations to remedy this unsatisfactory situation 
are discussed in~\cite{Kawamura:2012cr}.

The NNLO predictions for charm electro-production obtained for the two variants of the fit are shown in figure~\ref{fig:hera} for the 
kinematics of the HERA collider experiments. The two variants of the fit are not very different at large $Q^2$ which illustrates 
the moderate potential of the data~\cite{Aaron:2009af,Aaron:2009jy,Aaron:2011gp} to separate between different 
shapes of the NNLO term in the heavy-quark Wilson coefficients.  However, at small $Q^2$ the predictions diverge therefore 
the combined H1 and ZEUS  data may provide an additional constraint on this term and can improve the
accuracy of the NNLO determination of $m_c$. 

In the previous analysis~\cite{Alekhin:2010sv,Alekhin:2011jq}, values of the charm quark \msbar\ mass 
of $m_c(m_c)= 1.26\, \pm 0.09 ({\rm exp})\, \pm 0.11 ({\rm th})$ at NLO and $m_c(m_c)= 1.01\, \pm 0.09 ({\rm exp})\, \pm 0.03 ({\rm th})$ at 
NNLO$_\text{approx}$ was obtained. While the NLO results of the present and the previous analyses are consistent, the present result at NNLO$_\text{approx}$ is significantly larger. 
This shift in the central value is due to several sources.
First, in the present study we use a significantly improved theory of heavy-quark production in DIS. 
As mentioned above, we reconstruct the three-loop Wilson coefficient
$c_{2,i}^{(2)}$ based on all available information in various kinematic limits~\cite{Kawamura:2012cr}, 
in contrast to earlier studies~\cite{Alekhin:2010sv,Alekhin:2011jq} 
which relied only on the threshold approximation $c_{2,g}^{\,(2)\,{\rm thr}}$. 
Also, the analysis of~\cite{Alekhin:2010sv,Alekhin:2011jq}
was performed as a variant of the ABKM09 global analysis~\cite{Alekhin:2009ni} of world DIS data, 
to which the HERA experiments, at that time, only contributed the inclusive data of~\cite{Adloff:2000qk,Chekanov:2001qu}.
The present determination of $m_c(m_c)$ is based on the ABM11 fit~\cite{Alekhin:2012ig}, 
which incorporates the new combination of the HERA run I inclusive DIS data 
from the H1 and ZEUS experiments~\cite{Aaron:2009aa} and to which 
we add the new precise data sets on DIS charm production~\cite{Aaron:2009jy,Aaron:2011gp,Aaron:2009af}
discussed at above.
\begin{table}[h]
\begin{center}
\footnotesize
\renewcommand{\arraystretch}{1.3}
\begin{tabular}{|c|c|p{8.0cm}|}
\hline
$m_c(m_c)$ & reference & determination method 
\\ \hline
1.261 $\pm$ 0.016 & NARISON~\cite{Narison:2011xe} 
& QCD sum rules for vector current correlator 
\\ \hline
1.278 $\pm$ 0.009 & BODENSTEIN~\cite{Bodenstein:2011ma} 
&  QCD sum rules for vector current correlator
\\ \hline
1.28 $^{+0.07}_{-0.06}$ & LASCHKA~\cite{Laschka:2011zr} 
& lattice QCD; charmonium spectrum
\\ \hline
1.196 $\pm$ 0.059 $\pm$ 0.050 & AUBERT~\cite{Aubert:2009qda} 
&  inclusive spectra in semileptonic $B$-decays 
\\ \hline
1.28 $\pm$ 0.04  & BLOSSIER~\cite{Blossier:2010cr} 
& lattice QCD ($n_f =2$); hadron spectrum
\\ \hline
1.273 $\pm$ 0.006 & MCNEILE~\cite{McNeile:2010ji} 
& lattice QCD ($n_f$ = 2+1); pseudo-scalar current correlator
\\ \hline
1.279 $\pm$ 0.013 & CHETYRKIN~\cite{Chetyrkin:2009fv} 
& $e^+e^- \to c\bar{c}$ cross-section and QCD sum rules.
\\ \hline 
1.25 $\pm$ 0.04 & SIGNER~\cite{Signer:2008da} 
& non-relativistic QCD sum rules and $e^+e^- \to c \bar{c}$ cross-section near threshold.
\\ \hline
1.295 $\pm$ 0.015 & BOUGHEZAL~\cite{Boughezal:2006px} 
& $e^+e^- \to c \bar{c}$ cross-section. 
\\ \hline
1.24 $\pm$ 0.09 & BUCHMULLER~\cite{Buchmuller:2005zv} 
& global fit to inclusive $B$-decay spectra.
\\ \hline
1.224 $\pm$ 0.017 $\pm$ 0.054 &  HOANG~\cite{Hoang:2005zw} 
& global fit to inclusive $B$-decay data. 
\\ \hline
\end{tabular}
\caption{
  \small
  \label{tab:mc-compilation}
  Individual results of the charm-quark mass determinations used to compute the world
  average in the \msbar\ scheme~\cite{Beringer:2012}.
}
\normalsize
\end{center}
\end{table}

The values of $m_c(m_c)$ in equations~(\ref{eq:mcres-nlo}) and (\ref{eq:mcres-nnlo})
determined at NLO and at NNLO$_\text{approx}$ 
are close to the world average of $m_c(m_c)=1.275\pm 0.025$ GeV as evaluated by PDG~\cite{Beringer:2012}
and illustrated in figure~\ref{we_vs_all}.
The individual charm-quark mass determination used in the 
averaging procedure\footnote{Note that the PDG converts all results to $m_c(m_c)$ 
using scheme transformation~(\ref{eq:match}) at two loops in QCD together with
the value $\alpha_s(m_c) = 0.38 \pm 0.03$.}
are summarised in table~\ref{tab:mc-compilation}. 
They are based on different theoretical methods but are essentially limited 
to two approaches only.
They have either been obtained non-perturbatively with the help of lattice QCD 
simulations with a given number of dynamical fermions.
Or, in case perturbative QCD predictions have been applied, 
the extractions of $m_c(m_c)$ listed in table~\ref{tab:mc-compilation} are
limited to processes with time-like kinematics,
i.e., cross-section data from $e^+e^-$-collisions exposed to QCD sum rule analyses
or data on $B$-decays. 
The latter method carries also an intrinsic dependence 
on the uncertainty of the $b$-quark mass, while the former, 
i.e., the QCD sum rule analyses fix the value of $\alpha_s(M_Z)$ 
to the world average including the associated very small error{\footnote{
The current world average of $\alpha_s(M_Z)$ is the result of an arithmetic 
average of high precision determinations at least to NNLO
which are only marginally compatible within their quoted
errors, see e.g.,~\cite{Beringer:2012,Alekhin:2012ig}.}.
It has been shown~\cite{Dehnadi:2011gc}, 
that the systematic shift of $m_c(m_c)$ due to the value of $\alpha_s(M_Z)$ 
in QCD sum rule analyses is quite sizable.

\section{Conclusions}
The high precision measurements of charm quark production in DIS at HERA offer an attractive possibility 
to extract the charm-quark mass in the theoretically 
well-founded \msbar\ scheme. 
The resulting experimental precision of the present determination based on DIS data 
is substantially improved with respect to previous analyses.
It is now compatible with the theoretical uncertainty when comparing 
to the QCD predictions at NLO and it is significantly smaller than in case of the approximate NNLO QCD
predictions. The latter suffer from missing information on the three-loop Wilson coefficients 
at small-$x$ and small values of $Q^2$ and imply an additional theoretical uncertainty on $m_c$ of 100 MeV.

In comparison to the measurements used to define the world average up to now, our results 
add complementary information from scattering processes with
space-like kinematics and provide an important test of the QCD dynamics.
The kinematic range covered by the DIS data allows for the extraction 
of $m_c$ in the \msbar\ scheme well within the regime of validity of perturbative QCD. 
This analysis accounts for 
the full correlation of the dependence on $m_c$ with other non-perturbative parameters, 
most prominently the value of strong coupling constant $\alpha_s(M_Z)$ and the gluon PDF.
Future improvements on the accuracy of the $m_c(m_c)$ extractions from DIS data rely mostly
on the theoretical progress for the three-loop Wilson coefficients 
especially at small $x$. 
Also the experimental uncertainty can still be 
reduced by combining the charm production 
data from the H1 and the ZEUS collaborations. 

In summary, the charm-quark mass extractions from DIS data may challenge the accuracy 
of QCD sum rules analyses once the uncertainties on all non-perturbative parameters are accounted for on equal footing.

\subsection*{Acknowledgements}
This work has been supported in part by Helmholtz Gemeinschaft under contract VH-HA-101 ({\it Alliance Physics at the Terascale}), VH-NG-401 ({\it Young Investigator group "Physics of gluons and heavy quarks"}), by the Deutsche Forschungsgemeinschaft in 
Sonderforschungs\-be\-reich/Transregio~9 and 
by the European Commission through contract PITN-GA-2010-264564 ({\it LHCPhenoNet}).

{\footnotesize

\begin{thebibliography}{10}

\bibitem{Bigi:1994em}
I.~I. Bigi, M.~A. Shifman, N.~Uraltsev, and A.~Vainshtein,
\newblock Phys.Rev. {\bf D50}, 2234 (1994), hep-ph/9402360.

\bibitem{Beringer:2012}
Particle Data Group, J.~Beringer {\em et~al.},
\newblock Phys. Rev. {\bf D86}, 010001 (2012).

\bibitem{Alekhin:2010sv}
S.~Alekhin and S.~Moch,
\newblock Phys.Lett. {\bf B699}, 345 (2011), arXiv:1011.5790.

\bibitem{Aaron:2009jy}
H1 Collaboration, F.~Aaron {\em et~al.},
\newblock Phys.Lett. {\bf B686}, 91 (2009), arXiv:0911.3989.

\bibitem{Aaron:2011gp}
H1 Collaboration, F.~Aaron {\em et~al.},
\newblock Eur.Phys.J. {\bf C71}, 1769 (2011), arXiv:1106.1028.

\bibitem{Aaron:2009af}
H1 Collaboration, F.~Aaron {\em et~al.},
\newblock Eur.Phys.J. {\bf C65}, 89 (2010), arXiv:0907.2643.

\bibitem{Alekhin:2012ig}
S.~Alekhin, J.~Bl{\"u}mlein, and S.~Moch,
\newblock (2012), arXiv:1202.2281.

\bibitem{Alekhin:2011jq}
S.~Alekhin and S.~Moch,
\newblock (2011), arXiv:1107.0469.

\bibitem{Laenen:1992zk}
E.~Laenen, S.~Riemersma, J.~Smith, and W.~van Neerven,
\newblock Nucl.Phys. {\bf B392}, 162 (1993).

\bibitem{Riemersma:1994hv}
S.~Riemersma, J.~Smith, and W.~van Neerven,
\newblock Phys.Lett. {\bf B347}, 143 (1995), hep-ph/9411431.

\bibitem{Buza:1995ie}
M.~Buza {\em et~al.},
\newblock Nucl.Phys. {\bf B472}, 611 (1996), hep-ph/9601302.

\bibitem{Bierenbaum:2007qe}
I.~Bierenbaum, J.~Bl{\"u}mlein, and S.~Klein,
\newblock Nucl.Phys. {\bf B780}, 40 (2007), hep-ph/0703285.

\bibitem{Buza:1996wv}
M.~Buza, Y.~Matiounine, J.~Smith, and W.~van Neerven,
\newblock Eur.Phys.J. {\bf C1}, 301 (1998), hep-ph/9612398.

\bibitem{Alekhin:2009ni}
S.~Alekhin, J.~Bl{\"u}mlein, S.~Klein, and S.~Moch,
\newblock Phys.Rev. {\bf D81}, 014032 (2010), arXiv:0908.2766.

\bibitem{Forte:2010ta}
S.~Forte, E.~Laenen, P.~Nason, and J.~Rojo,
\newblock Nucl.Phys. {\bf B834}, 116 (2010), arXiv:1001.2312.

\bibitem{Kawamura:2012cr}
H.~Kawamura, N.~L. Presti, S.~Moch, and A.~Vogt,
\newblock (2012), arXiv:1205.5727.

\bibitem{Gray:1990yh}
N.~Gray, D.~J. Broadhurst, W.~Grafe, and K.~Schilcher,
\newblock Z.Phys. {\bf C48}, 673 (1990).

\bibitem{Chetyrkin:1999qi}
K.~Chetyrkin and M.~Steinhauser,
\newblock Nucl.Phys. {\bf B573}, 617 (2000), hep-ph/9911434.

\bibitem{Melnikov:2000qh}
K.~Melnikov and T.~v. Ritbergen,
\newblock Phys.Lett. {\bf B482}, 99 (2000), hep-ph/9912391.

\bibitem{Langenfeld:2009wd}
U.~Langenfeld, S.~Moch, and P.~Uwer,
\newblock Phys.Rev. {\bf D80}, 054009 (2009), arXiv:0906.5273.

\bibitem{Aliev:2010zk}
M.~Aliev {\em et~al.},
\newblock Comput.Phys.Commun. {\bf 182}, 1034 (2011), arXiv:1007.1327.

\bibitem{Vermaseren:1997fq}
J.~Vermaseren, S.~Larin, and T.~van Ritbergen,
\newblock Phys.Lett. {\bf B405}, 327 (1997), hep-ph/9703284.

\bibitem{Chetyrkin:1997dh}
K.~Chetyrkin,
\newblock Phys.Lett. {\bf B404}, 161 (1997), hep-ph/9703278.

\bibitem{Harris:1995tu}
B.~Harris and J.~Smith,
\newblock Nucl.Phys. {\bf B452}, 109 (1995), hep-ph/9503484.

\bibitem{Laenen:1998kp}
E.~Laenen and S.~Moch,
\newblock Phys.Rev. {\bf D59}, 034027 (1999), hep-ph/9809550.

\bibitem{Alekhin:2008hc}
S.~Alekhin and S.~Moch,
\newblock Phys.Lett. {\bf B672}, 166 (2009), arXiv:0811.1412.

\bibitem{Bierenbaum:2008yu}
I.~Bierenbaum, J.~Bl{\"u}mlein, S.~Klein, and C.~Schneider,
\newblock Nucl.Phys. {\bf B803}, 1 (2008), arXiv:0803.0273.

\bibitem{Bierenbaum:2009mv}
I.~Bierenbaum, J.~Bl{\"u}mlein, and S.~Klein,
\newblock Nucl.Phys. {\bf B820}, 417 (2009), arXiv:0904.3563.

\bibitem{Bierenbaum:2009zt}
I.~Bierenbaum, J.~Bl{\"u}mlein, and S.~Klein,
\newblock Phys.Lett. {\bf B672}, 401 (2009), arXiv:0901.0669.

\bibitem{Ablinger:2010ty}
J.~Ablinger {\em et~al.},
\newblock Nucl.Phys. {\bf B844}, 26 (2011), arXiv:1008.3347.

\bibitem{Catani:1990eg}
S.~Catani, M.~Ciafaloni, and F.~Hautmann,
\newblock Nucl.Phys. {\bf B366}, 135 (1991).

\bibitem{h1frag}
H1 Collaboration, F.~Aaron {\em et~al.},
\newblock Eur.Phys.J. {\bf C59}, 589 (2009), arXiv:0808.1003.

\bibitem{Kartvelishvili:1977pi}
V.~Kartvelishvili, A.~Likhoded, and V.~Petrov,
\newblock Phys.Lett. {\bf B78}, 615 (1978).

\bibitem{ptkink_aleph}
ALEPH Collaboration, R.~Barate {\em et~al.},
\newblock Phys. Rep. {\bf 294}, 1 (1998).

\bibitem{ptkink_delphi}
DELPHI Collaboration, P.~Abreu {\em et~al.},
\newblock Z.Phys. {\bf C73}, 11 (1996).

\bibitem{ptkink_markII}
MARCII Collaboration, G.~S. Abrams {\em et~al.},
\newblock Phys. Rev. Lett. {\bf 64}, 1334 (1990).

\bibitem{ptkink_pluto1}
PLUTO Collaboration, C.~Berger {\em et~al.},
\newblock Phys. Lett. {\bf B82}, 449 (1979).

\bibitem{ptkink_pluto2}
PLUTO Collaboration, C.~Berger {\em et~al.},
\newblock Phys. Lett. {\bf B97}, 459 (1980).

\bibitem{ptkink_tasso}
TASSO Collaboration, R.~Brandelik {\em et~al.},
\newblock Phys. Lett. {\bf B94} (1980).

\bibitem{lohrmann}
E.~Lohrmann,
\newblock (2011), arXiv:1112.3757.

\bibitem{Adloff:2000qk}
H1 Collaboration, C.~Adloff {\em et~al.},
\newblock Eur.Phys.J. {\bf C21}, 33 (2001), hep-ex/0012053.

\bibitem{Chekanov:2001qu}
ZEUS Collaboration, S.~Chekanov {\em et~al.},
\newblock Eur.Phys.J. {\bf C21}, 443 (2001), hep-ex/0105090.

\bibitem{Aaron:2009aa}
H1 and ZEUS Collaboration, F.~Aaron {\em et~al.},
\newblock JHEP {\bf 1001}, 109 (2010), arXiv:0911.0884.

\bibitem{Narison:2011xe}
S.~Narison,
\newblock Phys.Lett. {\bf B706}, 412 (2012), arXiv:1105.2922.

\bibitem{Bodenstein:2011ma}
S.~Bodenstein {\em et~al.},
\newblock Phys.Rev. {\bf D83}, 074014 (2011), arXiv:1102.3835.

\bibitem{Laschka:2011zr}
A.~Laschka, N.~Kaiser, and W.~Weise,
\newblock Phys.Rev. {\bf D83}, 094002 (2011), arXiv:1102.0945.

\bibitem{Aubert:2009qda}
BABAR Collaboration, B.~Aubert {\em et~al.},
\newblock Phys.Rev. {\bf D81}, 032003 (2010), arXiv:0908.0415.

\bibitem{Blossier:2010cr}
ETM Collaboration, B.~Blossier {\em et~al.},
\newblock Phys.Rev. {\bf D82}, 114513 (2010), arXiv:1010.3659.

\bibitem{McNeile:2010ji}
C.~McNeile {\em et~al.},
\newblock Phys.Rev. {\bf D82}, 034512 (2010), arXiv:1004.4285.

\bibitem{Chetyrkin:2009fv}
K.~Chetyrkin {\em et~al.},
\newblock Phys.Rev. {\bf D80}, 074010 (2009), arXiv:0907.2110.

\bibitem{Signer:2008da}
A.~Signer,
\newblock Phys.Lett. {\bf B672}, 333 (2009), arXiv:0810.1152.

\bibitem{Boughezal:2006px}
R.~Boughezal, M.~Czakon, and T.~Schutzmeier,
\newblock Phys.Rev. {\bf D74}, 074006 (2006), hep-ph/0605023.

\bibitem{Buchmuller:2005zv}
O.~Buchm{\"u}ller and H.~Fl{\"a}cher,
\newblock Phys.Rev. {\bf D73}, 073008 (2006), hep-ph/0507253.

\bibitem{Hoang:2005zw}
A.~H. Hoang and A.~V. Manohar,
\newblock Phys.Lett. {\bf B633}, 526 (2006), hep-ph/0509195.

\bibitem{Dehnadi:2011gc}
B.~Dehnadi, A.~H. Hoang, V.~Mateu, and S.~Zebarjad,
\newblock (2011), arXiv:1102.2264.

\end{thebibliography}

}

\newpage
\begin{figure}[t]
\center
  \includegraphics[width=0.9\textwidth]{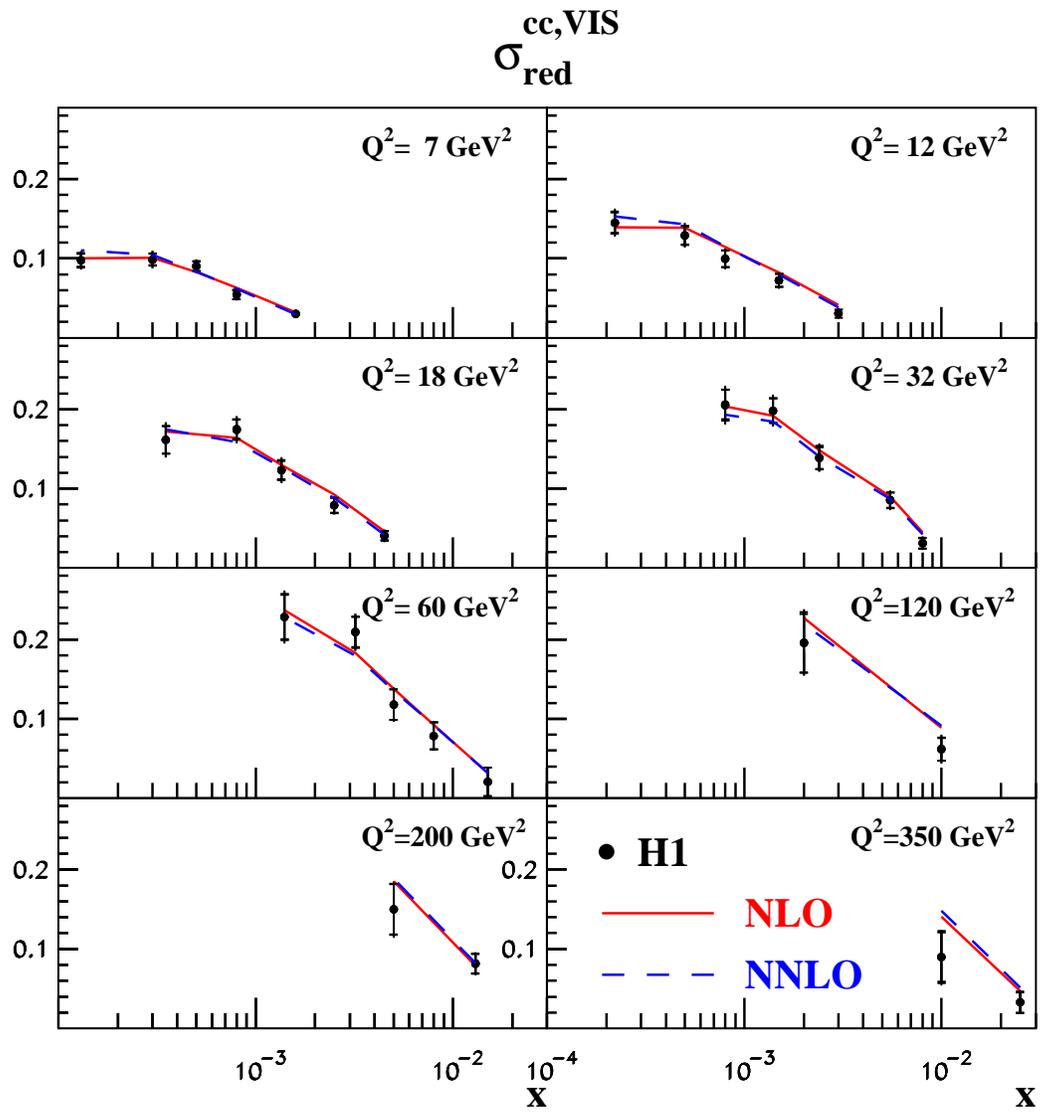}
\setlength{\unitlength}{1cm}
\caption{
The visible reduced charm quark production cross section, 
as determined from the measurement of $D^*$ meson production by the H1 
collaboration in comparison to the predicted cross sections 
using the NLO (solid line) and NNLO (dashed line) variants of this analysis.
}
\label{fig:h1sig} 
\end{figure}

\begin{figure}[hhh]
\center
  \includegraphics[width=0.9\textwidth]{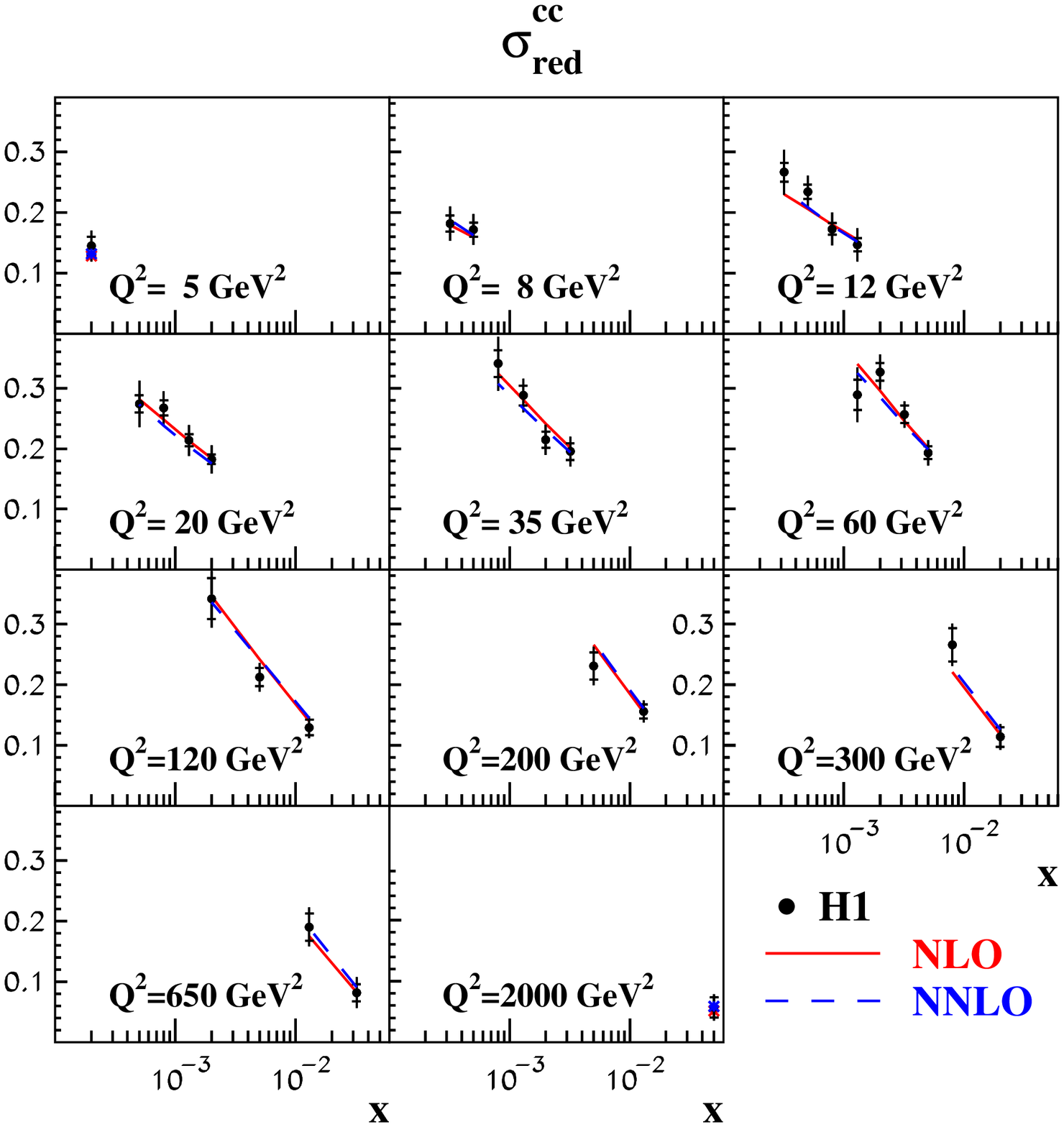}
\setlength{\unitlength}{1cm}
\caption{
  The reduced charm quark production cross section as measured by 
  the H1 collaboration using the vertex information of inclusive track production
  in comparison to  the predicted cross sections using the NLO (solid line) and NNLO
  (dashed line) variants of this analysis.
}
\label{fig:h1f2} 
\end{figure}

\begin{figure}[hhh]
\center
\epsfig{file=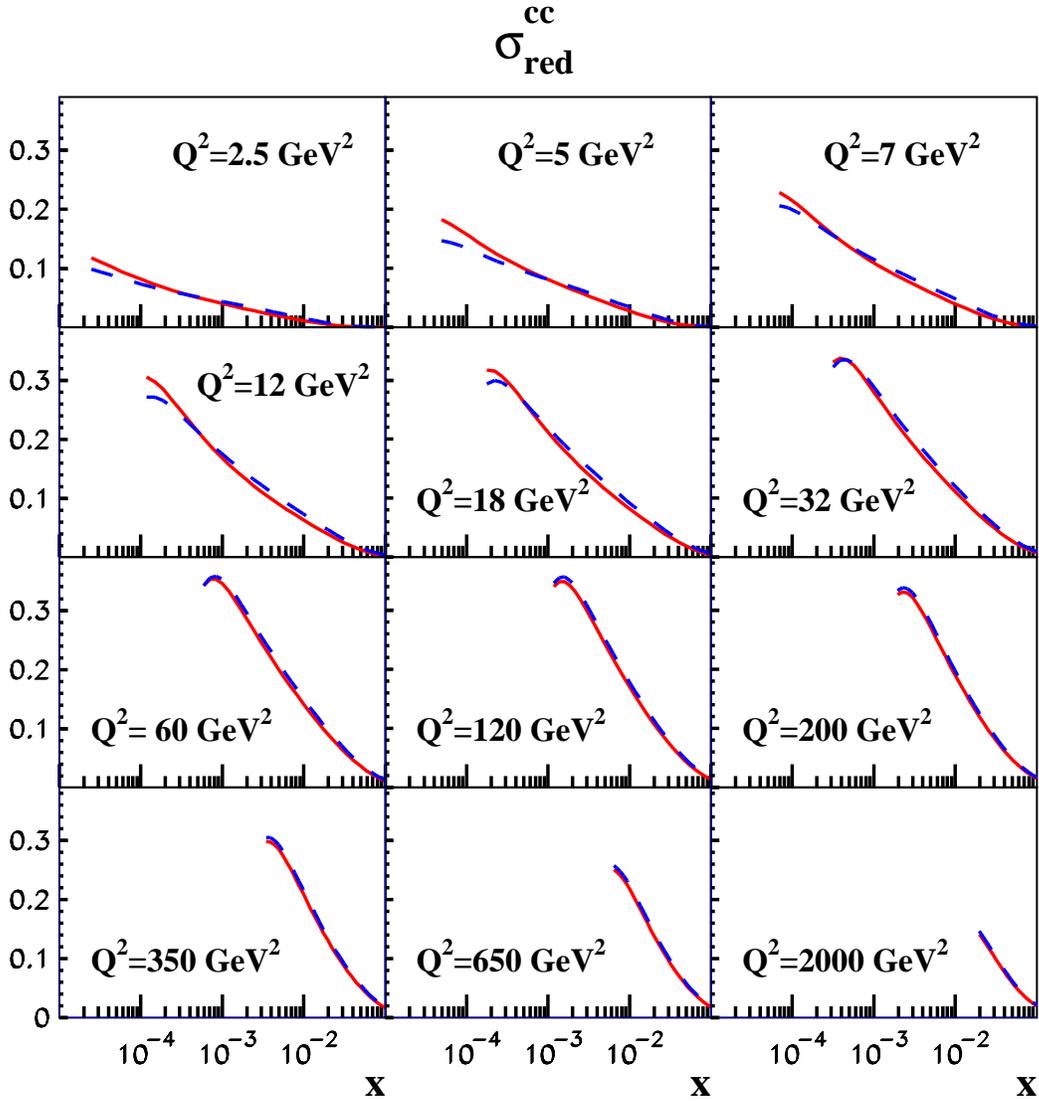,width=0.9\textwidth}
\setlength{\unitlength}{1cm}
\caption{
  The NNLO predictions for the reduced cross section of the charm electro-production as a function of $x$ for 
  different $Q^2$, for kinematics of the HERA collider experiments. 
  The solid line represents the variant of 
  the fit using $d_N=-0.6$ and $m_c(m_c)=1.36~{\rm GeV}$. The result corresponding to the fit 
  using $d_N=0$ and $m_c(m_c)=1.27~~{\rm GeV}$ is shown by a dashed line.
}
\label{fig:hera} 
\end{figure}

\begin{figure}[hhh]
\center
\epsfig{file=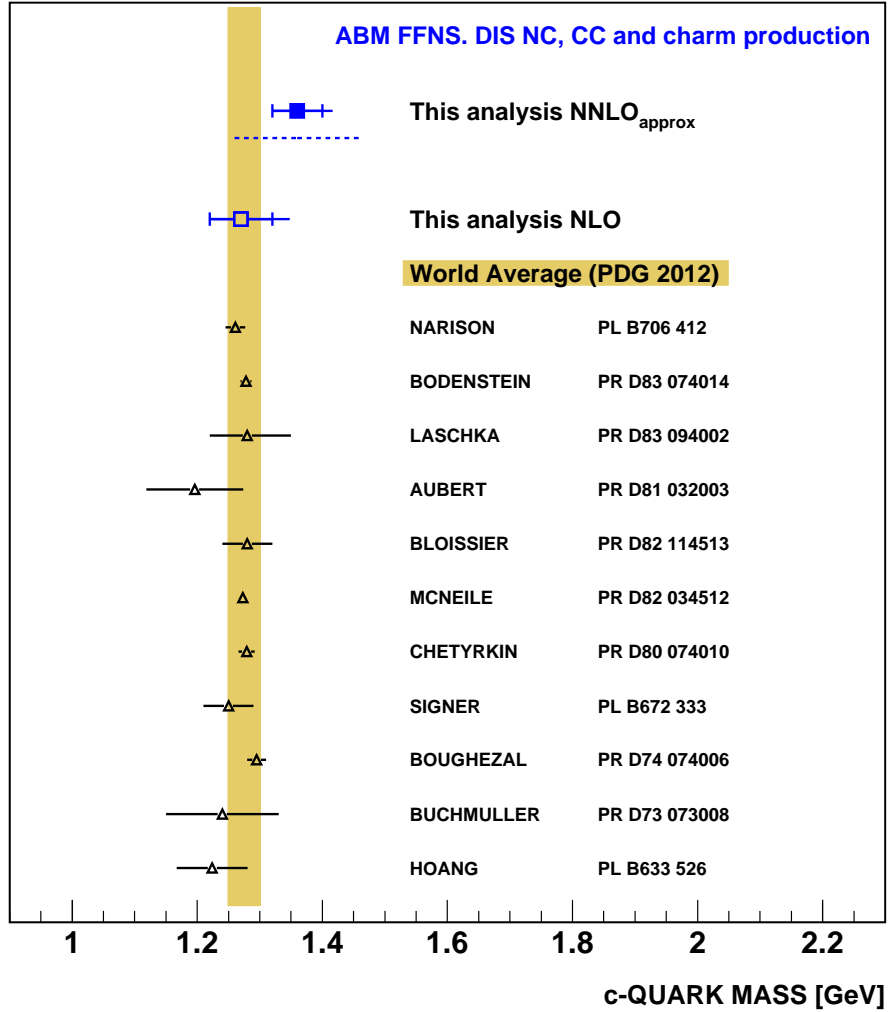,width=0.9\textwidth}
\setlength{\unitlength}{1cm}
\caption{
The charm quark \msbar\ mass as determined at NLO (open square) and at NNLO$_\text{approx}$ 
(filled square) using DIS data including the measurements of open charm
production at H1. 
The inner (outer) error bars represent experimental (total) uncertainties. 
To obtain the total uncertainty, the experimental and the theory uncertainties 
arising due to variation of the scales are added in quadrature. 
For comparison, the world average evaluated by the PDG~\cite{Beringer:2012} 
is shown by the shaded band. 
The measurements entering the world average determination are represented by
open triangles.
}
\label{we_vs_all} 
\end{figure}
\end{document}